\documentclass[aps,prb,reprint,groupedaddress]{revtex4-1}

\usepackage{graphicx}

\begin{document}

\title{Pyroelectric detection of spontaneous polarization
in magnetite thin films}

\author{R. Takahashi}
\email[]{rtaka@issp.u-tokyo.ac.jp}
\author{H. Misumi}
\author{M. Lippmaa}
\affiliation{Institute for Solid State Physics, University of Tokyo,
5-1-5, Kashiwanoha, Kashiwa, Chiba 277-8581, Japan}

\date{\today}

\begin{abstract}

We have investigated the spontaneous polarization in Fe$_3$O$_4$ thin
films by using dynamic and static pyroelectric measurements. The magnetic
and dielectric behavior of Fe$_3$O$_4$ thin films grown on
Nb:SrTiO$_3$(001) substrates was consistent with bulk crystals. The
well-known metal-insulator (Verwey) transition was observed at
120~K. The appearance of a pyroelectric response in the Fe$_3$O$_4$
thin films just below the Verwey temperature shows that spontaneous polarization
appeared in Fe$_3$O$_4$ at the charge ordering transition temperature.
The polar state characteristics are consistent with bond- and site- centered
charge ordering of Fe$^{2+}$ and Fe$^{3+}$ ions sharing the octahedral $B$ sites.
The pyroelectric response in Fe$_3$O$_4$ thin films was dependent on
the dielectric constant. Quasi-static pyroelectric measurement of
Pd/Fe$_3$O$_4$/Nb:SrTiO$_3$ junctions showed that magnetite has a very large
pyroelectric coefficient of 735~nC/cm$^{2}$K at 60~K. 

\end{abstract}


\maketitle

\section{INTRODUCTION}

Magnetite (Fe$_3$O$_4$) is a common magnetic ferrite that has an inverse
spinel structure with Fe$^{3+}$ ions occupying the tetrahedrally
coordinated $A$ sites and an equal number of Fe$^{2+}$ and Fe$^{3+}$ ions
sharing the octahedral $B$ sites. Exchange interactions between the different
iron sites are antiferromagnetic with the $A$-$B$ sublattice exchange
being dominant. This leads to ferrimagnetic spin ordering with a magnetic
moment per formula unit (f.u.) close to 4.05~$\mu_{\mathrm B}$ and a high
Curie temperature of 860~K.\cite{walz_jp, opel_jpd, monti_prb}
The magnetic properties of magnetite make it a useful material
for spintronic applications, such as tunnel
junctions \cite{seneor_apl, zaag_mmm} and spin injection
devices.\cite{wada_apl, yamanaka_nl} Another unique feature of Fe$_3$O$_4$
is the well-known metal-insulator Verwey transition at 120~K.\cite{verwey_n}
At room temperature, magnetite is metallic, since electrons can hop
within the $B$-site lattice between the Fe$^{2+}$ and Fe$^{3+}$ ions.
In contrast, the charges of Fe$^{2+}$ and Fe$^{3+}$
ions become ordered below 120~K and Fe$_3$O$_4$ crystals become
insulating. A remarkable feature of the insulating Fe$_3$O$_4$ phase is the
appearance of ferroelectricity. Fe$_3$O$_4$ is thus not only a prototype
multiferroic material with both spontaneous magnetization and dielectric
polarization, but also a rare ferroelectric crystal that appears in nature.
Most other common ferroelectric crystals, such as BaTiO$_3$,
Pb(Ti,Zr)O$_3$, etc., were artificially designed and synthesized.
Studies of the ferroelectric state in Fe$_3$O$_4$ started
with the magneto-electric measurements of Rado et al., which 
indicated the presence of a polar state in Fe$_3$O$_4$ crystals at
4.2~K.\cite{rado_prb_1975, rado_prb_1977} Subsequent works by Siratori et al.
found that the magneto-electric response from Fe$_3$O$_4$ crystals was tunable
by an electric field at 77~K.\cite{siratori_jpsj} Finally, Kato et al. succeeded in the
observation of ferroelectric switching at 4.2~K and reported that the
ferroelectric polarization along the $a$- and $c$-axis were 4.8~$\mu$C/cm$^2$
and 1.5~$\mu$C/cm$^2$, respectively.\cite{kato_jpsj, kato_jmmm} Measuring
ferroelectric polarization of magnetite at higher temperatures is hampered by
the gradual increase of conductivity as the Verwey temperature is approached.
The temperature dependence of  polarization in Fe$_3$O$_4$
has therefore been studied by measuring the pyroelectric response.
\cite{miyamoto_jpsj_1986, inase_jpsj} Recent reports on ferroelectric
Fe$_3$O$_4$ bulk crystals and thin films agree with the
earlier results, showing that hysteresis loops can be measured even with
a conventional ferroelectric tester, albeit only at temperatures
below 40~K.\cite{alexe_am, ziese_jpcm, schrettle_prb}

The origin of the spontaneous polarization cannot be explained by the
conventional mechanism of displacive ferroelectricity. The Fe$_3$O$_4$
low-temperature crystal structure belongs to a centrosymmetric monoclinic
symmetry group ($\it C_{c}$),\cite{iizumi_ac} which would normally
preclude the existence of spontaneous polarization.
However, recent theoretical work by Brink et al. has provided an
explanation for a possible origin of ferroelectricity in Fe$_3$O$_4$.
\cite{brink_jp,efremov_nm} Below the Verwey transition at
120~K, a regular arrangement of the $B$-site Fe$^{2+}$ and Fe$^{3+}$ ions in
an inverse spinel structure results in a charge-ordered pattern.
The $B$-sites form a pyrochlore lattice consisting of
corner-sharing tetrahedra, where an alternating pattern of Fe$^{2+}$ and
Fe$^{3+}$ ions forms along the monoclinic $b$-axis direction of Fe$_3$O$_4$.
The charge ordering results in an alternation of short and
long Fe-Fe bonds. The coexistence of bond-centered and charge-centered
charge ordering induces an electronic polarization along the monoclinic
$b$-axis. \cite{brink_jp,yamauchi_prb, fukushima_jpsj}  This model is supported by structural analysis of powder
diffraction refinements \cite{wright_prl, blasco_prb} and resonant X-ray scattering studies.
\cite{joly_prb, lorenzo_prl} Furthermore, Senn et al. have recently succeeded
in accurate structural analysis by high-energy X-ray diffraction
from a single-domain Fe$_3$O$_4$ sample and fully determined the
low-temperature superstructure of a Fe$_3$O$_4$ crystal.\cite{senn_n, senn_prb}
They showed that the charge ordering results in three-site distortions
that induce substantial off-center atomic displacements and
couple to the resulting large dielectric polarization.
Similar electronic ferroelectricity induced by charge ordering is also
known for LuFe$_2$O$_4$ \cite{ikeda_n, ikeda_jpsj} and
Pr(Sr$_{0.1}$Ca$_{0.9}$)$_2$Mn$_2$O$_7$.\cite{tokunaga_nm}
It is clear that in such materials there is a relationship between
the pattern of charge ordering and polarization.
However, there is still a discrepancy in the Fe$_3$O$_4$
experimental results, since spontaneous polarization
has only been observed well below the
Verwey transition point, 
\cite{rado_prb_1975, rado_prb_1977, siratori_jpsj, kato_jpsj, kato_jmmm, 
miyamoto_jpsj_1986, inase_jpsj, alexe_am, ziese_jpcm, schrettle_prb}
while X-ray and neutron diffraction studies suggest that
the charge-ordered state responsible for the ferroelectricity
appears at the Verwey transition and is unchanged upon further cooling.
\cite{wright_prl, joly_prb, senn_n, senn_prb} One possible reason for this
discrepancy is the high leakage current of Fe$_3$O$_4$ crystals
just below the charge-ordering temperature of 120~K. 
      
In order to investigate the relationship between the electronic
polarization and the Verwey transition, we have studied by dynamic and
static pyroelectric detection the temperature dependence of spontaneous
polarization in Fe$_3$O$_4$ thin films grown on Nb:SrTiO$_3$(001)
substrates. The pyroelectric measurement was performed at zero applied bias,
which means that the results are insensitive to temperature-dependent
resistivity changes close to the Verwey temperature.\cite{chynoweth_jap, lines,
bune_jap, takahashi_apl, takahashi_jap} The dynamic pyroelectric
response of a Fe$_3$O$_4$ junction was used to study the relationship between
the spontaneous polarization and the Verwey transition. 
Quasi-static pyroelectric analysis was used to determine the absolute pyroelectric
coefficient of Pd/Fe$_3$O$_4$/Nb:SrTiO$_3$(001) junctions. 

\section{EXPERIMENT}

The Fe$_3$O$_4$ thin films were grown by pulsed laser deposition  
on 0.2$^\circ$ miscut Nb(0.05wt\%):SrTiO$_3$(001) substrates that had been
wet-etched in buffered NH$_4$F-HF to obtain a well-defined surface
termination.\cite{kawasaki_s, koster_apl} A polycrystalline Fe$_2$O$_3$
target was ablated with an excimer laser at a fluence of 3~J/cm$^2$ under an oxygen
background pressure of $1\times10^{-6}$~Torr. The ablation laser (KrF,
$\lambda=248$~nm) operated at 10~Hz. The Fe$_3$O$_4$ film thicknesses were
between 150 and 220~nm. The growth temperature was set at 400$^\circ$C
and the temperature control was done with an infrared laser
heater.\cite{ohashi_rsi} After growth,
the films were rapidly cooled below 200$^\circ$C in about 5 min in
order to suppress the oxidization of Fe and the formation of a secondary
hematite Fe$_2$O$_3$ phase.\cite{takahashi_cgd}

The basic structural analysis was done at room temperature by symmetric X-ray diffraction
and reciprocal space mapping. Magnetization of the Fe$_3$O$_4$ thin films was
measured in a superconducting quantum interference device (SQUID) magnetometer
at 5~K and 300~K. Raman spectroscopy was used for detecting structural
transitions below room temperature. A He-Ne laser (633~nm, 17~mW) was
focused onto a Fe$_3$O$_4$ film surface through an objective lens
($\times50$, N.A. = 0.5). The scattering spectra were collected by a
charge coupled device (CCD) detector (RAMASCOPE, Renishaw). The sample
temperature was controlled with a He flow cryostat (Microstat, Oxford
Instruments). 

For electrical measurements, a 100-nm-thick Pd top electrode was
deposited on the Fe$_3$O$_4$ film surface by electron beam evaporation
through a stencil mask with 1~mm diameter openings. Aluminum
wires were attached to the Pd top electrode pads with silver paste.
The sample was placed in a vacuum chamber and cooled using a two-stage
cryocooler. The sample temperature was controlled in the rage of 8 to 300~K
by thermal conduction from the cryocooler and a heater mounted on the sample
stage. The resistance was measured by two-point method using a
picoammeter (Keithley 487). The dielectric measurements
were performed with an impedance bridge (Agilent 4284A) at an excitation
voltage of 50~mV. 

The Chynoweth method was for dynamic pyroelectric
measurements.\cite{chynoweth_jap, lines, bune_jap, takahashi_apl, takahashi_jap}
Chopped light from a diode laser (1.31~$\mu$m, 130~mW) was focused on a Pd
top electrode pad, resulting in a modulation of the Fe$_3$O$_4$ capacitor
temperature and the generation of a pyroelectric current.
The laser chopping was achieved by modulating the diode laser current
with an optical power risetime of $\approx3\mu$s. The sample
current was converted to a voltage signal with a current-voltage converter at a
transconductance of $10^8$~V/A and measured with a digital voltmeter or a
lock-in amplifier. For ferroelectric hysteresis measurements, a 20-nm-thick TiN
bottom electrode was inserted between a Fe$_3$O$_4$ film and
a SrTiO$_3$(001) substrate in order to promote charge screening
during ferroelectric switching. Details of the hysteresis
loop measurements can be found in Refs. \cite{takahashi_apl} and \cite{takahashi_jap}.
A quasi-static
pyroelectric measurement was used for measuring the absolute pyroelectric
coefficient by slowly heating and cooling a sample
in a temperature-stabilized probing chamber.\cite{bune_jap} The
pyroelectric current was measured with a picoammeter for Fe$_3$O$_4$
samples alternately heated and cooled at a constant rate, ranging from 1
to 6~K/min. The temperature of the film was monitored with a Si diode
mounted next to a Fe$_3$O$_4$ capacitor sample. 

\section{RESULTS AND DISCUSSION}

\begin{figure}
\includegraphics{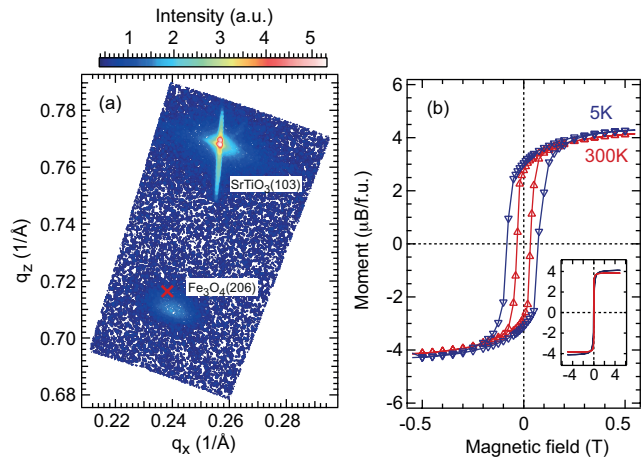}
\caption{(color online) (a) Reciprocal space map of a magnetite film
around the Nb:SrTiO$_3$(103) substrate reflection. The lattice parameter
of bulk Fe$_3$O$_4$ crystal is marked with a cross. (b) Magnetization curves
at 5~K and 300~K for Fe$_3$O$_4$ thin films. The saturated magnetization was
4~$\mu_{\mathrm B}$/f.u., consistent with the bulk Fe$_3$O$_4$ magnetization.
Wide-range magnetization curves measured at ($\bigtriangledown$) 5~K
and ($\bigtriangleup$) 300~K are show in the inset.}
\label{fig1}
\end{figure}

Fig.~\ref{fig1}(a) shows a reciprocal space map around
the Nb:SrTiO$_3$(103) reflection, indicating cube-on-cube growth of
a (001)-oriented Fe$_3$O$_4$ film on the Nb:SrTiO$_3$(001) substrate.
The in-plane and out-of-plane lattice parameters were 8.33~\AA\ 
and 8.44~\AA, consistent with the bulk lattice parameter of
$a$=8.396~\AA.\cite{JCPDS_mag}
As expected, the 150~nm-thick Fe$_3$O$_4$ film was almost fully
relaxed on the SrTiO$_3$(001) substrate, since the lattice mismatch
between Fe$_3$O$_4$ and SrTiO$_3$ is -7.5\%.\cite{JCPDS_STO, JCPDS_mag}

In general, Fe$_3$O$_4$ films are known to include antiphase boundaries (APBs)
caused by random nucleation at the initial growth stage of a spinel on a
perovskite substrate. The presence of such boundaries can influence the
film characteristics. For example, the magnetization of defect-rich films
remains unsaturated even in magnetic fields of 7~T.\cite{margulies_prb}
The magnetization loops of the Fe$_3$O$_4$ thin films used in this work 
are shown in Fig.~\ref{fig1}(b). 
The saturated magnetization was approximately 4~$\mu_{\mathrm B}$/f.u.,
matching the bulk crystal value.\cite{walz_jp, opel_jpd}
The high-field data, plotted in the inset of Fig.~\ref{fig1}(b), 
shows that the magnetization was fully saturated at fields below 2 T, 
suggesting that the density of APBs in our laser-ablated films was sufficiently low
not to influence the attainment of magnetic saturation at this film thickness. \cite{hibma_jap}  
Furthermore, the temperature dependence of magnetization (not shown) showed
that the Verwey transition temperature was 120K,
confirming that the Fe$_3$O$_4$ film composition was very close
to stoichiometric.\cite{aragon_prb}
Together with the structural analysis, the magnetic behavior showed that
bulk-equivalent samples were obtained.

\begin{figure}
\includegraphics{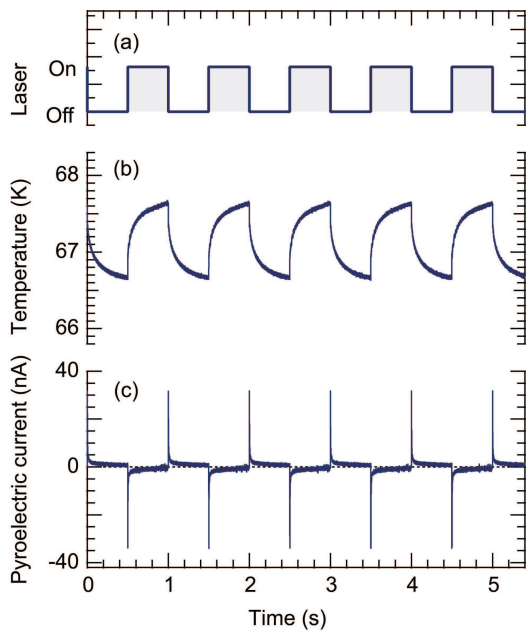}
\caption{(color online) Transient profiles of the heating laser power (a), 
thin film sample temperature (b) and the pyroelectric current (c) for a
Pd/Fe$_3$O$_4$/Nb:SrTiO$_3$ junction.}
\label{dynamic_pyro}
\end{figure}

Fig.~\ref{dynamic_pyro} shows the transient profiles of the heating
laser operation (a), sample temperature (b), and pyroelectric current (c)
for a Pd/ Fe$_3$O$_4$/ Nb:SrTiO$_3$ capacitor, measured at an ambient
temperature of 67~K and a laser pulse rate of 1~Hz with a 50\% duty cycle.
The sample temperature variation 
was measured with a Si diode and separately
calculated from the known static temperature dependence by measuring the
sample resistance variation during pulsed laser illumination.
The temperature change was exponential for
both heating and cooling phases of each measurement period.
For a laser chopping frequency of 1~Hz, the capacitor temperature
variation amplitude was estimated at 1~K,
resulting in the generation of negative and positive spike currents,
as shown in Fig.~\ref{dynamic_pyro}(c). The current spike is caused by the
temperature dependence of the spontaneous polarization in polar materials.
The measurement thus shows that spontaneous polarization existed in the
Fe$_3$O$_4$ film at the measurement temperature of 67~K.
  
  \begin{figure}
\includegraphics{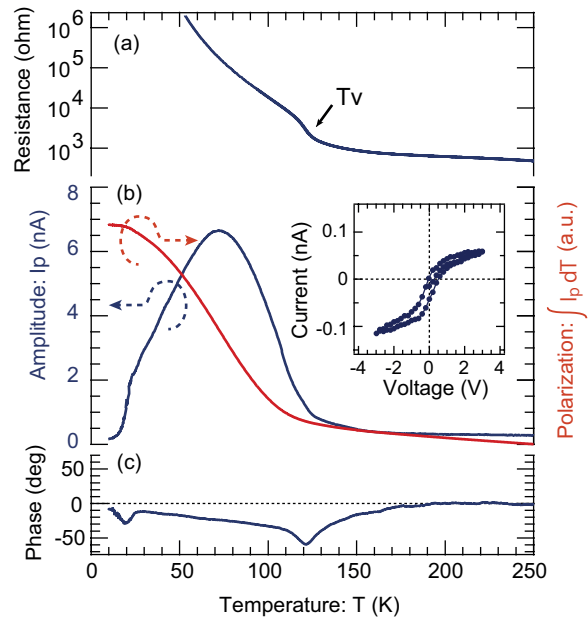}
\caption{(color online) Temperature dependence of resistance (a), dynamic
pyroelectric current amplitude (b) and phase (c) for a Pd/Fe$_3$O$_4$/Nb:SrTiO$_3$
junction. The Verwey transition is visible at 120~K. The right axis in (b)
corresponds to the integrated pyroelectric current, showing the temperature
dependence of polarization in the Fe$_3$O$_4$ film. The inset in (b) shows
a hysteresis loop for a Pd/Fe$_3$O$_4$/TiN junction measured at 9~K.
The polarization was switchable by an applied electric field,
proving that the films were ferroelectric.}
\label{pyro_t_nbsto}
\end{figure}

In order to investigate the relationship between the Verwey transition
and the appearance of spontaneous polarization in Fe$_3$O$_4$ films,
the temperature dependence of the resistance and the pyroelectric response
amplitude were measured for a Pd/Fe$_3$O$_4$/Nb:SrTiO$_3$ capacitor, as
shown in Fig.~\ref{pyro_t_nbsto}. Fig.~\ref{pyro_t_nbsto}(a) shows the
discontinuous resistance change at 120~K, corresponding to the bulk Verwey
transition temperature.\cite{verwey_n} Below the Verwey transition
temperature, the pyroelectric response increased rapidly with
decreasing sample temperature. As shown by the plot in Fig.~\ref{pyro_t_nbsto}(b),
the maximum pyroelectric signal amplitude was observed at 70~K,
below which the pyroelectric response was reduced. The pyroelectric
current is generally proportional to the differential of the ferroelectric
polarization. The temperature dependence of polarization was estimated
by integrating the pyroelectric current, with the assumption that the
temperature variation induced by the infrared laser was independent of
the measurement temperature. The inset of Fig.~\ref{pyro_t_nbsto}(b) shows
the hysteresis measurement result for a Pd/Fe$_3$O$_4$/TiN junction at
9~K. The pyroelectric current polarity was switched by
an applied electric field, indicating the presence of ferroelectricity in
the Pd/Fe$_3$O$_4$/TiN junction.

\begin{figure}
\includegraphics{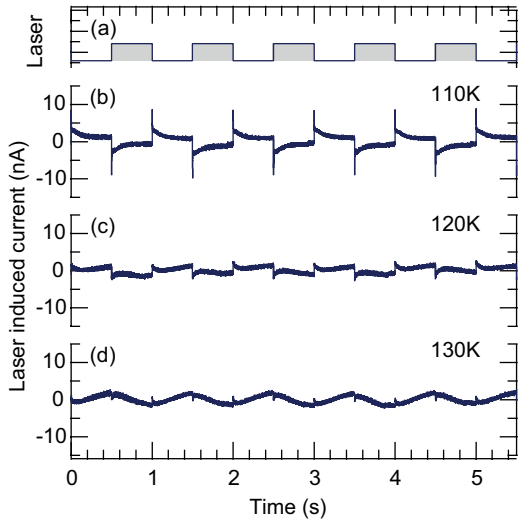}
\caption{(color online) Transient profiles of the heating laser power (a) 
and the pyroelectric current for a Pd/Fe$_3$O$_4$/Nb:SrTiO$_3$ junction at
110~K (b), 120~K (c), and 130~K (d). }
\label{transient_120K}
\end{figure}

Discontinuities in the pyroelectric response can be seen best in the
phase signal obtained from lock-in detection of the
sample current. Fig.~\ref{pyro_t_nbsto}(c) shows that there
were two peaks at 20~K and 120~K. The peak at 120~K is due to the Verwey
transition and is thus related to the charge ordering of Fe$^{2+}$ and
Fe$^{3+}$ ions. Fig.~\ref{transient_120K} shows the transient profiles
of laser power (a) and pyroelectric currents measured at 110~K (b), 120~K (c),
and 130~K (d). At 110~K and 120~K, positive and negative
spike currents are clearly visible, revealing the presence of spontaneous
polarization in the Fe$_3$O$_4$ thin film. In contrast, the
polarization-related current spikes disappear above the
Verwey transition temperature and only a contribution from the
temperature-dependent leak current through the sample capacitor remains,
resulting in the discontinuous modulation of pyroelectric
current phase at 120~K in Fig.~\ref{pyro_t_nbsto}(c).

\begin{figure}
\includegraphics{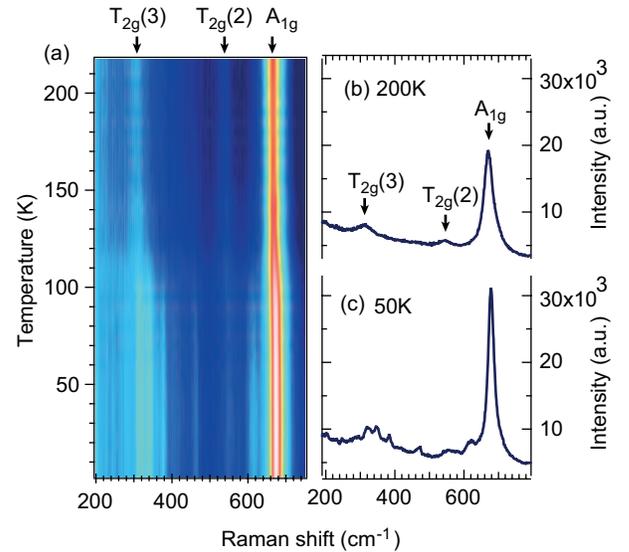}
\caption{(color online) (a) Temperature dependence of Raman spectra for a
Fe$_3$O$_4$ thin film. Individual Raman spectra measured at 200~K (b)
and 50~K (c).}
\label{raman}
\end{figure}

Raman measurements were performed on
Fe$_3$O$_4$/Nb:SrTiO$_3$ junctions between 10~K and 300~K
in order to look for possible crystal symmetry changes in the thin film
samples at low temperatures.
The temperature dependence of the Raman spectra is shown
in Fig.~\ref{raman}(a). A spectrum taken at 200~K, well above the
Verwey temperature, is shown in Fig.~\ref{raman}(b). The three main
features in the spectrum can be assigned to the A$_{1g}$,
T$_{2g}$(2), and T$_{2g}$(3) modes by comparing the Raman shifts
with literature data.\cite{verble_prb, phase_jap} These three peaks
did not show significant change between room temperature and the Verwey
transition point. At the Verwey transition temperature of 120~K, the crystal
structure of Fe$_3$O$_4$ changes from the high-temperature cubic phase to
the low-temperature monoclinic symmetry. The effect of the symmetry reduction
at the Verwey transition point can be seen in the splitting of the
T$_{2g}$(3) peak into three components and a large increase in the
A$_{1g}$ peak intensity. These changes can be seen in the comparison of
the 200~K and 50~K Raman spectra in Figs.~\ref{raman}(b) and \ref{raman}(c).
A discontinuity in the Raman spectra occurred at around 115~K,
consistent with the temperature dependence of the resistance shown in
Fig.~\ref{pyro_t_nbsto}(a).
The Raman measurements reveal only a single structural transition in the
10~K to 300~K temperature range, located at the Verwey temperature.
No other structural changes were seen at lower temperatures.
This observation is consistent with previous nuclear
magnetic resonance (NMR) \cite{kovtun_ssc, yanai_jpsj} and heat capacity
\cite{schrettle_prb} results for bulk Fe$_3$O$_4$ crystals. This
implies that the peak at 20~K in Fig.~\ref{pyro_t_nbsto}(c) is not due
to a structural transition. 

The pyroelectric current amplitude is strongly dependent on the heat capacity and the thermal
conductivity of a material and the sample capacitance. \cite{lines}
The heat capacity of Fe$_3$O$_4$ shows a gradual monotonic decrease
 below the Verwey transition temperature.\cite{schrettle_prb}
In contrast, the thermal conductivity of Fe$_3$O$_4$ shows a gradual monotonic increase
below the Verwey temperature down to about 25~K and an exponential drop at
lower temperatures.\cite{slack_pr}
This suggests that the temperature variation induced by periodic
heating with a chopped constant-power laser source may change
at around 25~K. However, due to the small thickness of the film,
the main contribution to the temperature modulation amplitude comes
from the 0.5-mm-thick Nb:SrTiO$_3$ substrate.
This suggests that the thermal conductivity change in Fe$_3$O$_4$ does not
have a significant effect on the pyroelectric current and is not the
cause for the phase jump in Fig.~\ref{pyro_t_nbsto}(b) at around 20~K.

\begin{figure}
\includegraphics{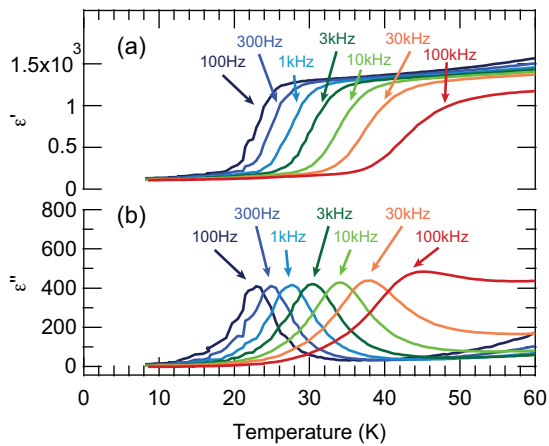}
\caption{(color online) Temperature dependence of (a) the dielectric
permittivity $\epsilon'$ and (b) loss $\epsilon''$ for a
Pd/Fe$_3$O$_4$/Nb:SrTiO$_3$ junction for several measurement frequencies.}
\label{dielectric}
\end{figure}

A clue to the origin of the phase signal feature at 20~K is offered
by the measurement of the device permittivity, shown in
Fig.~\ref{dielectric}, where the real and imaginary components of the
magnetite film permittivity are plotted as a function of temperature
for several measurement frequencies.
A strong dielectric dispersion is visible, with a 
drop of the dielectric constant at temperatures below 50~K,
consistent with previous reports.\cite{iwauchi_jpsj, kobayashi_jpsj,
alexe_am, ziese_jpcm, schrettle_prb}
The phase response shown in Fig.~\ref{pyro_t_nbsto}(c) was measured
at a chopping rate of 35~Hz. At this frequency, the drop of permittivity
occurs at around 20~K, as shown in Fig.~\ref{dielectric}. When a pyroelectric
measurement is performed at such a dielectric transition edge, the
laser-induced temperature change also modulates the device capacitance,
resulting in a slight asymmetry in the heating and cooling phase
pyroelectric current spikes. A lock-in current detector sees the asymmetry
as a slight phase shift, explaining the phase jump at 20~K in
Fig.~\ref{pyro_t_nbsto}(c).

\begin{figure}
\includegraphics{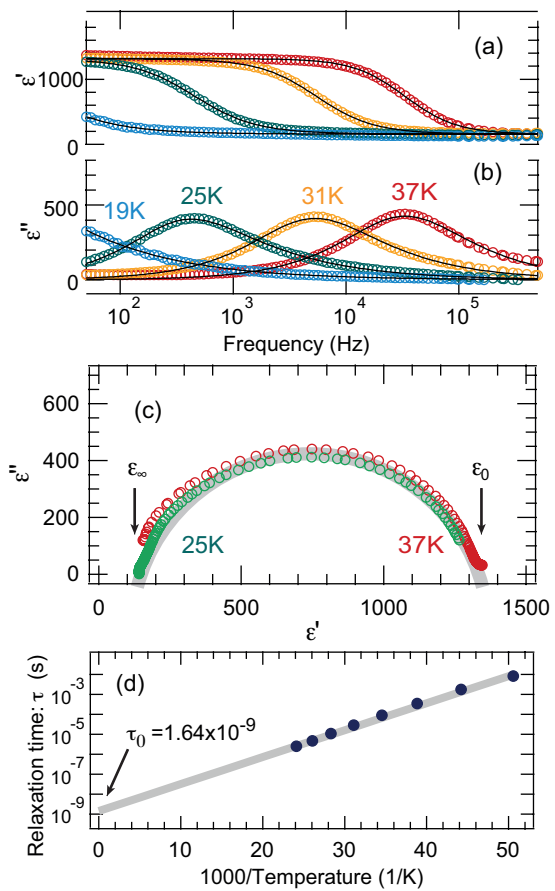}
\caption{(color online) Temperature dependence of (a) $\epsilon'$ and
(b) $\epsilon''$ for a Pd/Fe$_3$O$_4$/Nb:SrTiO$_3$ junction for several
measurement frequencies. The circles and solid lines represent the measured
data and the Havriliak-Negami fitting results, respectively. (c) A Cole-Cole
plot of $\epsilon'$ and $\epsilon''$, yielding estimates for $\epsilon_\infty=135$
at 25~K and $\epsilon_0=1340$ at 37~K. (d) Arrhenius plot of relaxation
times obtained from the Havriliak-Negami fitting. The activation
energy and prefactor were estimated at 26.8~meV and 1.64~ns.}
\label{debye}
\end{figure}

The dielectric response plots in Fig.~\ref{dielectric} show strong dispersion,
which is a common feature for order-disorder type ferroelectric
materials \cite{fu_prl} and
has been attributed in Fe$_3$O$_4$ to electron transfer between Fe$^{2+}$
and Fe$^{3+}$ ions.\cite{iwauchi_jpsj} A similar phenomenon is
known to occur in the electronic ferroelectric material,
LuFe$_2$O$_4$.\cite{ikeda_n, ikeda_jpsj} In general, the motion of a ferroelectric
domain boundary gives rise to the dispersion, indicating the presence of
ferroelectric domains and boundary motion. In order to investigate
the dielectric dispersion in Fe$_3$O$_4$ films, the frequency
dependence of $\epsilon'$ and $\epsilon''$ were measured at
19~K, 25~K, 31~K, and 37~K, and plotted in Figs.~\ref{debye}(a) and
\ref{debye}(b), showing a typical signature of relaxation. A stepwise
change in $\epsilon'$ is accompanied by a $\epsilon''$ peak.
At higher temperatures, the peak position of
$\epsilon''$ shifted to higher frequencies, implying that the
relaxation time $\tau$ became smaller, since $\tau$=1/(2$\pi\times$frequency).
The empirical Havriliak-Negami function \cite{havriliak_p} was used to 
quantitatively analyze the dielectric relaxation behavior.
\begin{equation}
\epsilon'-i\epsilon''=\epsilon_\infty +
  \frac{\epsilon_0-\epsilon_\infty}{[1+(i\omega\tau)^{1-\alpha}]^\beta}, 
\label{eq:hn}
\end{equation}
where $\epsilon_\infty$ and $\epsilon_0$ are the high- and
low-frequency dielectric constants, respectively, and $\alpha$ and
$\beta$ represent the broadening and asymmetry factors of the curves.
The permittivities estimated from the Cole-Cole plots in Fig.~\ref{debye}(c)
were $\epsilon_\infty=135$ and $\epsilon_0=1340$.
After substituting the estimated $\epsilon_\infty$ and
$\epsilon_0$ into the Eq.~\ref{eq:hn}, $\alpha$, $\beta$, and $\tau$ were
optimized. The fitting results are shown by 
solid lines in Figs.~\ref{debye}(a) and \ref{debye}(b). The Havriliak-Negami
function provides excellent fits for the observed dielectric dispersion.
Fig.~\ref{debye}(d) shows values of $\tau$ from the fitting results
in an Arrhenius representation.
\begin{equation}
\tau=\tau_0\exp\left(\frac{E_a}{k_{\mathrm B}T}\right),
\label{eq:acti}
\end{equation}
where $\tau_0$ is a prefactor, $k_{\mathrm B}T$ is the thermal energy,
and $E_a$ is the activation energy. 
The observed linear slope in Fig.~\ref{debye}(d) shows that the
dielectric relaxation is thermally activated with $E_a=26.8$~meV,
which is close to the reported values for bulk and film Fe$_3$O$_4$.
\cite{iwauchi_jpsj, kobayashi_jpsj, alexe_am, ziese_jpcm, schrettle_prb} 
Linear extrapolation yields a prefactor
value of $\tau_0=1.64$~ns, corresponding to a rather low attempt frequency
of less than 100~MHz. This value is much lower than the 280~GHz of a typical
electronic ferroelectric, LuFe$_2$O$_4$\cite{ikeda_n} and
would appear to show that much larger effective domain sizes are
involved.\cite{senn_n, senn_prb}

\begin{figure}
\includegraphics{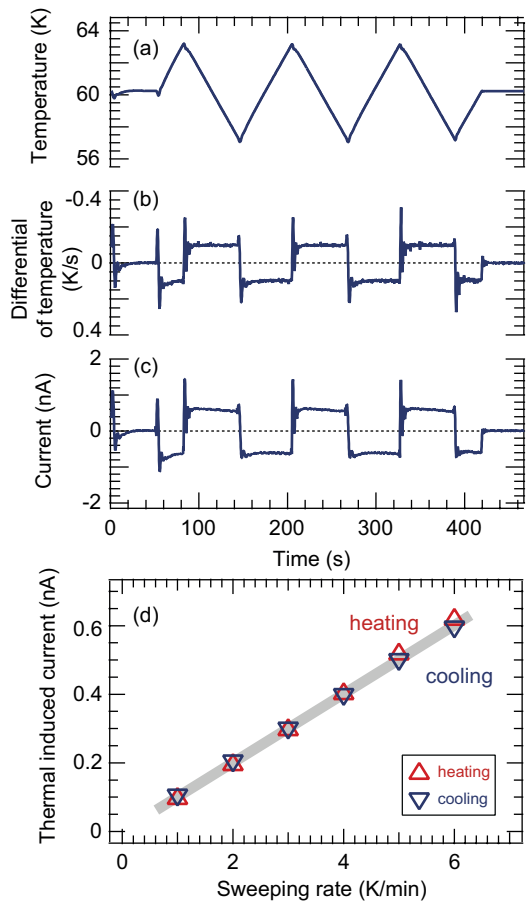}
\caption{(color online) Static pyroelectric response for
a Pd/Fe$_3$O$_4$/Nb:SrTiO$_3$ junction. (a) Time profile of the slow
heating and cooling cycles applied to the sample. (b) Differential
of the temperature cycle. (c) A constant pyroelectric current was observed
when the film was heated or cooled at a constant rate at an average
temperature of 60~K. (d) The effective pyroelectric
coefficient estimated from the slope of the thermally-induced current
was 735~nC/cm$^2$K.}
\label{static_pyro}
\end{figure}

The large dielectric constant above 50~K
plays an important role in Pd/Fe$_3$O$_4$/Nb:SrTiO$_3$ junctions
for obtaining the strong pyroelectric response shown in 
Figs.~\ref{dynamic_pyro} and \ref{pyro_t_nbsto}(b).
The maximum pyroelectric current in Fig.~\ref{pyro_t_nbsto}(b) is
approximately 7~nA and independent of the polarity of the poling bias.
In order to evaluate the absolute value of the pyroelectric coefficient,
a quasi-static pyroelectric measurement without the use of laser
illumination was performed for a Pd/Fe$_3$O$_4$/Nb:SrTiO$_3$ junction at 60~K.
This method can be used to eliminate possible photo-induced currents
that may affect the dynamic pyroelectric measurements.\cite{takahashi_jap}
Figs. \ref{static_pyro}(a-c) show the slow sample temperature sweep,
the measured sample temperature gradient, and the
observed pyroelectric current, respectively. The temperature sweep rate
was 6~K/min. At a constant average temperature of 60~K, there was no
generated current, as expected for dark conditions. During the heating slope,
a constant negative pyroelectric current was generated and a corresponding
positive pyroelectric current was observed for the cooling slope. This
can be verified by comparing Figs.~\ref{static_pyro}(b) and \ref{static_pyro}(c).
The transient spikes in the differential of the sample temperature
in Fig.~\ref{static_pyro}(b) are caused by the finite settling time of the
temperature controller and corresponding pyroelectric current spikes can also be
seen in Fig.~\ref{static_pyro}(c). The nonideal temperature controller
performance thus served as a convenient way to check that
the observed slow pyroelectric response is consistent with the faster
response measured in the dynamic measurements with chopped optical heating.

The static pyroelectric current was proportional to the temperature sweep rate,
as shown in Fig.~\ref{static_pyro}(d). From the slope of a linear fit of the
data points, the effective pyroelectric coefficient at 60~K was estimated
to be 735~nC/cm$^2$K. This value is comparable to the well-known large
pyroelectric coefficient of Pb(Sc,Ta)O$_3$ films, 600~nC/cm$^2$K,\cite{watton_f}
and much larger than, e.g., in PbTiO$_3$/MgO films, 20~nC/cm$^2$K.\cite{iijima_jap}
We note that the pyroelectric coefficient involves a piezoelectric contribution
due to a difference in the thermal expansion coefficients of the Fe$_3$O$_4$ film
and the Nb:SrTiO$_3$ substrate.\cite{bune_jap, lines}

\section{CONCLUSION}

We have demonstrated dynamic and static pyroelectric measurements
of Fe$_3$O$_4$ thin film capacitors. 
It was shown that the polar state in Fe$_3$O$_4$ does indeed appear at the
Verwey transition point.
The relaxor behavior at low temperatures is consistent with
the spontaneous polarization being induced by bond- and site-centered charge
ordering of Fe$^{2+}$ and Fe$^{3+}$ ions sharing the octahedral $B$ sites.

The dynamic pyroelectric response was influenced by the sharp change
of the dielectric permittivity in the 20~K to 50~K range. The dielectric
transition at $\approx20$~K was confirmed not to be caused by a structural
transition. The slow characteristic time scale
of the relaxor response at $\approx1$~ns appears to indicate that the
charge-ordered relaxor domains involve a large number of iron sites.
Static pyroelectric measurement were used to show that magnetite
has a very large pyroelectric coefficient of 735~nC/cm$^2$K at 60~K.

\begin{acknowledgments}

The authors would like to thank the Asahi Glass Foundation, Futaba
Electronics Memorial Foundation, Foundation for Interaction in Science \& Technology
 and the Murata Science Foundation for funding. 

\end{acknowledgments}

\bibliography{ref.bib}

\providecommand{\noopsort}[1]{}\providecommand{\singleletter}[1]{#1}%
\begin{thebibliography}{57}%
\makeatletter
\providecommand \@ifxundefined [1]{%
 \@ifx{#1\undefined}
}%
\providecommand \@ifnum [1]{%
 \ifnum #1\expandafter \@firstoftwo
 \else \expandafter \@secondoftwo
 \fi
}%
\providecommand \@ifx [1]{%
 \ifx #1\expandafter \@firstoftwo
 \else \expandafter \@secondoftwo
 \fi
}%
\providecommand \natexlab [1]{#1}%
\providecommand \enquote  [1]{``#1''}%
\providecommand \bibnamefont  [1]{#1}%
\providecommand \bibfnamefont [1]{#1}%
\providecommand \citenamefont [1]{#1}%
\providecommand \href@noop [0]{\@secondoftwo}%
\providecommand \href [0]{\begingroup \@sanitize@url \@href}%
\providecommand \@href[1]{\@@startlink{#1}\@@href}%
\providecommand \@@href[1]{\endgroup#1\@@endlink}%
\providecommand \@sanitize@url [0]{\catcode `\\12\catcode `\$12\catcode
  `\&12\catcode `\#12\catcode `\^12\catcode `\_12\catcode `\%12\relax}%
\providecommand \@@startlink[1]{}%
\providecommand \@@endlink[0]{}%
\providecommand \url  [0]{\begingroup\@sanitize@url \@url }%
\providecommand \@url [1]{\endgroup\@href {#1}{\urlprefix }}%
\providecommand \urlprefix  [0]{URL }%
\providecommand \Eprint [0]{\href }%
\@ifxundefined \urlstyle {%
  \providecommand \doi  [0]{\begingroup \@sanitize@url \@doi}%
  \providecommand \@doi [1]{\endgroup \@@startlink {\doibase
  #1}doi:\discretionary {}{}{}#1\@@endlink }%
}{%
  \providecommand \doi  [0]{doi:\discretionary{}{}{}\begingroup
  \urlstyle{rm}\Url }%
}%
\providecommand \doibase [0]{http://dx.doi.org/}%
\providecommand \Doi [0]{\begingroup \@sanitize@url \@Doi }%
\providecommand \@Doi  [1]{\endgroup\@@startlink{\doibase#1}\@@Doi}%
\providecommand \@@Doi [1]{#1\@@endlink}%
\providecommand \selectlanguage [0]{\@gobble}%
\providecommand \bibinfo  [0]{\@secondoftwo}%
\providecommand \bibfield  [0]{\@secondoftwo}%
\providecommand \translation [1]{[#1]}%
\providecommand \BibitemOpen [0]{}%
\providecommand \bibitemStop [0]{}%
\providecommand \bibitemNoStop [0]{.\EOS\space}%
\providecommand \EOS [0]{\spacefactor3000\relax}%
\providecommand \BibitemShut  [1]{\csname bibitem#1\endcsname}%
\bibitem [{\citenamefont {Walz}(2002)}]{walz_jp}%
  \BibitemOpen
  \bibfield  {author} {\bibinfo {author} {\bibfnamefont {F.}~\bibnamefont
  {Walz}},\ }\href@noop {} {\bibfield  {journal} {\bibinfo  {journal} {J.
  Phys.:Condens. Matter.},\ }\textbf {\bibinfo {volume} {14}},\ \bibinfo
  {pages} {R285} (\bibinfo {year} {2002})}\BibitemShut {NoStop}%
\bibitem [{\citenamefont {Opel}(2012)}]{opel_jpd}%
  \BibitemOpen
  \bibfield  {author} {\bibinfo {author} {\bibfnamefont {M.}~\bibnamefont
  {Opel}},\ }\href@noop {} {\bibfield  {journal} {\bibinfo  {journal} {J. Phys.
  D: Appl. Phys.},\ }\textbf {\bibinfo {volume} {45}},\ \bibinfo {pages}
  {033001} (\bibinfo {year} {2012})}\BibitemShut {NoStop}%
\bibitem [{\citenamefont {Monti}\ \emph {et~al.}(2012)\citenamefont {Monti},
  \citenamefont {Santos}, \citenamefont {Mascaraque}, \citenamefont {de~la
  Fuente}, \citenamefont {Nino}, \citenamefont {Mentes}, \citenamefont
  {Locatelli}, \citenamefont {MacCarty}, \citenamefont {Marco},\ and\
  \citenamefont {de~la Figuera}}]{monti_prb}%
  \BibitemOpen
  \bibfield  {author} {\bibinfo {author} {\bibfnamefont {M.}~\bibnamefont
  {Monti}}, \bibinfo {author} {\bibfnamefont {B.}~\bibnamefont {Santos}},
  \bibinfo {author} {\bibfnamefont {A.}~\bibnamefont {Mascaraque}}, \bibinfo
  {author} {\bibfnamefont {O.~R.}\ \bibnamefont {de~la Fuente}}, \bibinfo
  {author} {\bibfnamefont {M.~A.}\ \bibnamefont {Nino}}, \bibinfo {author}
  {\bibfnamefont {T.~O.}\ \bibnamefont {Mentes}}, \bibinfo {author}
  {\bibfnamefont {A.}~\bibnamefont {Locatelli}}, \bibinfo {author}
  {\bibfnamefont {K.~F.}\ \bibnamefont {MacCarty}}, \bibinfo {author}
  {\bibfnamefont {J.~F.}\ \bibnamefont {Marco}}, \ and\ \bibinfo {author}
  {\bibfnamefont {J.}~\bibnamefont {de~la Figuera}},\ }\href@noop {} {\bibfield
   {journal} {\bibinfo  {journal} {Phys. Rev. B},\ }\textbf {\bibinfo {volume}
  {85}},\ \bibinfo {pages} {020404} (\bibinfo {year} {2012})}\BibitemShut
  {NoStop}%
\bibitem [{\citenamefont {Seneor}\ \emph {et~al.}(1999)\citenamefont {Seneor},
  \citenamefont {Fert}, \citenamefont {Maurice}, \citenamefont {Montaigne},
  \citenamefont {Petroff},\ and\ \citenamefont {Vaures}}]{seneor_apl}%
  \BibitemOpen
  \bibfield  {author} {\bibinfo {author} {\bibfnamefont {P.}~\bibnamefont
  {Seneor}}, \bibinfo {author} {\bibfnamefont {A.}~\bibnamefont {Fert}},
  \bibinfo {author} {\bibfnamefont {J.~L.}\ \bibnamefont {Maurice}}, \bibinfo
  {author} {\bibfnamefont {F.}~\bibnamefont {Montaigne}}, \bibinfo {author}
  {\bibfnamefont {F.}~\bibnamefont {Petroff}}, \ and\ \bibinfo {author}
  {\bibfnamefont {A.}~\bibnamefont {Vaures}},\ }\href@noop {} {\bibfield
  {journal} {\bibinfo  {journal} {Appl. Phys. Lett.},\ }\textbf {\bibinfo
  {volume} {74}},\ \bibinfo {pages} {4017} (\bibinfo {year}
  {1999})}\BibitemShut {NoStop}%
\bibitem [{\citenamefont {van~der Zaag}\ \emph {et~al.}(2000)\citenamefont
  {van~der Zaag}, \citenamefont {Bloemen}, \citenamefont {Gaines},
  \citenamefont {Wolf}, \citenamefont {van~der Heijden}, \citenamefont {van~de
  Veerdonk},\ and\ \citenamefont {de~Jonge}}]{zaag_mmm}%
  \BibitemOpen
  \bibfield  {author} {\bibinfo {author} {\bibfnamefont {P.~J.}\ \bibnamefont
  {van~der Zaag}}, \bibinfo {author} {\bibfnamefont {P.~J.~H.}\ \bibnamefont
  {Bloemen}}, \bibinfo {author} {\bibfnamefont {J.~M.}\ \bibnamefont {Gaines}},
  \bibinfo {author} {\bibfnamefont {R.~M.}\ \bibnamefont {Wolf}}, \bibinfo
  {author} {\bibfnamefont {P.~A.~A.}\ \bibnamefont {van~der Heijden}}, \bibinfo
  {author} {\bibfnamefont {R.~J.~M.}\ \bibnamefont {van~de Veerdonk}}, \ and\
  \bibinfo {author} {\bibfnamefont {W.~J.~M.}\ \bibnamefont {de~Jonge}},\
  }\href@noop {} {\bibfield  {journal} {\bibinfo  {journal} {J. Mag. Mag.
  Mater.},\ }\textbf {\bibinfo {volume} {211}},\ \bibinfo {pages} {301}
  (\bibinfo {year} {2000})}\BibitemShut {NoStop}%
\bibitem [{\citenamefont {Wada}\ \emph {et~al.}(2010)\citenamefont {Wada},
  \citenamefont {Watanabe}, \citenamefont {Shirahata}, \citenamefont {Itoh},
  \citenamefont {Yamaguchi},\ and\ \citenamefont {Taniyama}}]{wada_apl}%
  \BibitemOpen
  \bibfield  {author} {\bibinfo {author} {\bibfnamefont {E.}~\bibnamefont
  {Wada}}, \bibinfo {author} {\bibfnamefont {K.}~\bibnamefont {Watanabe}},
  \bibinfo {author} {\bibfnamefont {Y.}~\bibnamefont {Shirahata}}, \bibinfo
  {author} {\bibfnamefont {M.}~\bibnamefont {Itoh}}, \bibinfo {author}
  {\bibfnamefont {M.}~\bibnamefont {Yamaguchi}}, \ and\ \bibinfo {author}
  {\bibfnamefont {T.}~\bibnamefont {Taniyama}},\ }\href@noop {} {\bibfield
  {journal} {\bibinfo  {journal} {Appl. Phys. Lett.},\ }\textbf {\bibinfo
  {volume} {96}},\ \bibinfo {pages} {102510} (\bibinfo {year}
  {2010})}\BibitemShut {NoStop}%
\bibitem [{\citenamefont {Yamanaka}\ \emph {et~al.}(2011)\citenamefont
  {Yamanaka}, \citenamefont {Kanki}, \citenamefont {Kawai},\ and\ \citenamefont
  {Tanaka}}]{yamanaka_nl}%
  \BibitemOpen
  \bibfield  {author} {\bibinfo {author} {\bibfnamefont {S.}~\bibnamefont
  {Yamanaka}}, \bibinfo {author} {\bibfnamefont {T.}~\bibnamefont {Kanki}},
  \bibinfo {author} {\bibfnamefont {T.}~\bibnamefont {Kawai}}, \ and\ \bibinfo
  {author} {\bibfnamefont {H.}~\bibnamefont {Tanaka}},\ }\href@noop {}
  {\bibfield  {journal} {\bibinfo  {journal} {Nano Lett.},\ }\textbf {\bibinfo
  {volume} {11}},\ \bibinfo {pages} {343} (\bibinfo {year} {2011})}\BibitemShut
  {NoStop}%
\bibitem [{\citenamefont {Verwey}(1939)}]{verwey_n}%
  \BibitemOpen
  \bibfield  {author} {\bibinfo {author} {\bibfnamefont {E.~J.~W.}\
  \bibnamefont {Verwey}},\ }\href@noop {} {\bibfield  {journal} {\bibinfo
  {journal} {Nature},\ }\textbf {\bibinfo {volume} {144}},\ \bibinfo {pages}
  {327} (\bibinfo {year} {1939})}\BibitemShut {NoStop}%
\bibitem [{\citenamefont {Rado}\ and\ \citenamefont
  {Ferrari}(1975)}]{rado_prb_1975}%
  \BibitemOpen
  \bibfield  {author} {\bibinfo {author} {\bibfnamefont {G.~T.}\ \bibnamefont
  {Rado}}\ and\ \bibinfo {author} {\bibfnamefont {J.~M.}\ \bibnamefont
  {Ferrari}},\ }\href@noop {} {\bibfield  {journal} {\bibinfo  {journal} {Phys.
  Rev. B},\ }\textbf {\bibinfo {volume} {12}},\ \bibinfo {pages} {5166}
  (\bibinfo {year} {1975})}\BibitemShut {NoStop}%
\bibitem [{\citenamefont {Rado}\ and\ \citenamefont
  {Ferrari}(1977)}]{rado_prb_1977}%
  \BibitemOpen
  \bibfield  {author} {\bibinfo {author} {\bibfnamefont {G.~T.}\ \bibnamefont
  {Rado}}\ and\ \bibinfo {author} {\bibfnamefont {J.~M.}\ \bibnamefont
  {Ferrari}},\ }\href@noop {} {\bibfield  {journal} {\bibinfo  {journal} {Phys.
  Rev. B},\ }\textbf {\bibinfo {volume} {15}},\ \bibinfo {pages} {290}
  (\bibinfo {year} {1977})}\BibitemShut {NoStop}%
\bibitem [{\citenamefont {Siratori}\ \emph {et~al.}(1979)\citenamefont
  {Siratori}, \citenamefont {Kita}, \citenamefont {Kaji}, \citenamefont
  {Tasaki}, \citenamefont {Kimura}, \citenamefont {Shindo},\ and\ \citenamefont
  {Kohn}}]{siratori_jpsj}%
  \BibitemOpen
  \bibfield  {author} {\bibinfo {author} {\bibfnamefont {K.}~\bibnamefont
  {Siratori}}, \bibinfo {author} {\bibfnamefont {E.}~\bibnamefont {Kita}},
  \bibinfo {author} {\bibfnamefont {G.}~\bibnamefont {Kaji}}, \bibinfo {author}
  {\bibfnamefont {A.}~\bibnamefont {Tasaki}}, \bibinfo {author} {\bibfnamefont
  {S.}~\bibnamefont {Kimura}}, \bibinfo {author} {\bibfnamefont
  {I.}~\bibnamefont {Shindo}}, \ and\ \bibinfo {author} {\bibfnamefont
  {K.}~\bibnamefont {Kohn}},\ }\href@noop {} {\bibfield  {journal} {\bibinfo
  {journal} {J. Phys. Soc. Jpn.},\ }\textbf {\bibinfo {volume} {47}},\ \bibinfo
  {pages} {1779} (\bibinfo {year} {1979})}\BibitemShut {NoStop}%
\bibitem [{\citenamefont {Kato}\ and\ \citenamefont {Iida}(1982)}]{kato_jpsj}%
  \BibitemOpen
  \bibfield  {author} {\bibinfo {author} {\bibfnamefont {K.}~\bibnamefont
  {Kato}}\ and\ \bibinfo {author} {\bibfnamefont {S.}~\bibnamefont {Iida}},\
  }\href@noop {} {\bibfield  {journal} {\bibinfo  {journal} {J. Phys. Soc.
  Jpn.},\ }\textbf {\bibinfo {volume} {51}},\ \bibinfo {pages} {1335} (\bibinfo
  {year} {1982})}\BibitemShut {NoStop}%
\bibitem [{\citenamefont {Kato}\ \emph {et~al.}(1983)\citenamefont {Kato},
  \citenamefont {Iida}, \citenamefont {Yanai},\ and\ \citenamefont
  {Mizushima}}]{kato_jmmm}%
  \BibitemOpen
  \bibfield  {author} {\bibinfo {author} {\bibfnamefont {K.}~\bibnamefont
  {Kato}}, \bibinfo {author} {\bibfnamefont {S.}~\bibnamefont {Iida}}, \bibinfo
  {author} {\bibfnamefont {K.}~\bibnamefont {Yanai}}, \ and\ \bibinfo {author}
  {\bibfnamefont {K.}~\bibnamefont {Mizushima}},\ }\href@noop {} {\bibfield
  {journal} {\bibinfo  {journal} {J. Mag. Mag. Mater.},\ }\textbf {\bibinfo
  {volume} {31-34}},\ \bibinfo {pages} {783} (\bibinfo {year}
  {1983})}\BibitemShut {NoStop}%
\bibitem [{\citenamefont {Miyamoto}\ \emph {et~al.}(1986)\citenamefont
  {Miyamoto}, \citenamefont {Kobayashi},\ and\ \citenamefont
  {Chikazumi}}]{miyamoto_jpsj_1986}%
  \BibitemOpen
  \bibfield  {author} {\bibinfo {author} {\bibfnamefont {Y.}~\bibnamefont
  {Miyamoto}}, \bibinfo {author} {\bibfnamefont {M.}~\bibnamefont {Kobayashi}},
  \ and\ \bibinfo {author} {\bibfnamefont {S.}~\bibnamefont {Chikazumi}},\
  }\href@noop {} {\bibfield  {journal} {\bibinfo  {journal} {J. Phys. Soc.
  Jpn.},\ }\textbf {\bibinfo {volume} {55}},\ \bibinfo {pages} {660} (\bibinfo
  {year} {1986})}\BibitemShut {NoStop}%
\bibitem [{\citenamefont {Inase}\ and\ \citenamefont
  {Miyamoto}(1987)}]{inase_jpsj}%
  \BibitemOpen
  \bibfield  {author} {\bibinfo {author} {\bibfnamefont {T.}~\bibnamefont
  {Inase}}\ and\ \bibinfo {author} {\bibfnamefont {Y.}~\bibnamefont
  {Miyamoto}},\ }\href@noop {} {\bibfield  {journal} {\bibinfo  {journal} {J.
  Phys. Soc. Jpn.},\ }\textbf {\bibinfo {volume} {56}},\ \bibinfo {pages}
  {3683} (\bibinfo {year} {1987})}\BibitemShut {NoStop}%
\bibitem [{\citenamefont {Alexe}\ \emph {et~al.}(2009)\citenamefont {Alexe},
  \citenamefont {Ziese}, \citenamefont {Hesse}, \citenamefont {Esquinazi},
  \citenamefont {Yamauchi}, \citenamefont {Fukushima}, \citenamefont {Picozzi},
  ,\ and\ \citenamefont {Gosele}}]{alexe_am}%
  \BibitemOpen
  \bibfield  {author} {\bibinfo {author} {\bibfnamefont {M.}~\bibnamefont
  {Alexe}}, \bibinfo {author} {\bibfnamefont {M.}~\bibnamefont {Ziese}},
  \bibinfo {author} {\bibfnamefont {D.}~\bibnamefont {Hesse}}, \bibinfo
  {author} {\bibfnamefont {P.}~\bibnamefont {Esquinazi}}, \bibinfo {author}
  {\bibfnamefont {K.}~\bibnamefont {Yamauchi}}, \bibinfo {author}
  {\bibfnamefont {T.}~\bibnamefont {Fukushima}}, \bibinfo {author}
  {\bibfnamefont {S.}~\bibnamefont {Picozzi}}, , \ and\ \bibinfo {author}
  {\bibfnamefont {U.}~\bibnamefont {Gosele}},\ }\href@noop {} {\bibfield
  {journal} {\bibinfo  {journal} {Adv. Mater.},\ }\textbf {\bibinfo {volume}
  {21}},\ \bibinfo {pages} {4452} (\bibinfo {year} {2009})}\BibitemShut
  {NoStop}%
\bibitem [{\citenamefont {Ziese}\ \emph {et~al.}(2012)\citenamefont {Ziese},
  \citenamefont {Esquinazi}, \citenamefont {Pantel}, \citenamefont {Alexe},
  \citenamefont {Nemes},\ and\ \citenamefont {Gercia-Hernandez}}]{ziese_jpcm}%
  \BibitemOpen
  \bibfield  {author} {\bibinfo {author} {\bibfnamefont {M.}~\bibnamefont
  {Ziese}}, \bibinfo {author} {\bibfnamefont {P.~D.}\ \bibnamefont
  {Esquinazi}}, \bibinfo {author} {\bibfnamefont {D.}~\bibnamefont {Pantel}},
  \bibinfo {author} {\bibfnamefont {M.}~\bibnamefont {Alexe}}, \bibinfo
  {author} {\bibfnamefont {N.~M.}\ \bibnamefont {Nemes}}, \ and\ \bibinfo
  {author} {\bibfnamefont {M.}~\bibnamefont {Gercia-Hernandez}},\ }\href@noop
  {} {\bibfield  {journal} {\bibinfo  {journal} {J. Phys.:Condens. Matter},\
  }\textbf {\bibinfo {volume} {24}},\ \bibinfo {pages} {086007} (\bibinfo
  {year} {2012})}\BibitemShut {NoStop}%
\bibitem [{\citenamefont {Schrettle}\ \emph {et~al.}(2011)\citenamefont
  {Schrettle}, \citenamefont {Krohns}, \citenamefont {Lunkenheimer},
  \citenamefont {Brabers},\ and\ \citenamefont {Loidl}}]{schrettle_prb}%
  \BibitemOpen
  \bibfield  {author} {\bibinfo {author} {\bibfnamefont {F.}~\bibnamefont
  {Schrettle}}, \bibinfo {author} {\bibfnamefont {S.}~\bibnamefont {Krohns}},
  \bibinfo {author} {\bibfnamefont {P.}~\bibnamefont {Lunkenheimer}}, \bibinfo
  {author} {\bibfnamefont {V.~A.~M.}\ \bibnamefont {Brabers}}, \ and\ \bibinfo
  {author} {\bibfnamefont {A.}~\bibnamefont {Loidl}},\ }\href@noop {}
  {\bibfield  {journal} {\bibinfo  {journal} {Phys. Rev. B},\ }\textbf
  {\bibinfo {volume} {83}},\ \bibinfo {pages} {195109} (\bibinfo {year}
  {2011})}\BibitemShut {NoStop}%
\bibitem [{\citenamefont {Iizumi}\ \emph {et~al.}(1982)\citenamefont {Iizumi},
  \citenamefont {Koetzle}, \citenamefont {Shirane}, \citenamefont {Chikazumi},
  \citenamefont {Matsui},\ and\ \citenamefont {Todo}}]{iizumi_ac}%
  \BibitemOpen
  \bibfield  {author} {\bibinfo {author} {\bibfnamefont {M.}~\bibnamefont
  {Iizumi}}, \bibinfo {author} {\bibfnamefont {T.~F.}\ \bibnamefont {Koetzle}},
  \bibinfo {author} {\bibfnamefont {G.}~\bibnamefont {Shirane}}, \bibinfo
  {author} {\bibfnamefont {S.}~\bibnamefont {Chikazumi}}, \bibinfo {author}
  {\bibfnamefont {M.}~\bibnamefont {Matsui}}, \ and\ \bibinfo {author}
  {\bibfnamefont {S.}~\bibnamefont {Todo}},\ }\href@noop {} {\bibfield
  {journal} {\bibinfo  {journal} {Acta. Cryst.},\ }\textbf {\bibinfo {volume}
  {B38}},\ \bibinfo {pages} {2121} (\bibinfo {year} {1982})}\BibitemShut
  {NoStop}%
\bibitem [{\citenamefont {Brink}\ and\ \citenamefont
  {Khomskii}(2008)}]{brink_jp}%
  \BibitemOpen
  \bibfield  {author} {\bibinfo {author} {\bibfnamefont {J.}~\bibnamefont
  {Brink}}\ and\ \bibinfo {author} {\bibfnamefont {D.~I.}\ \bibnamefont
  {Khomskii}},\ }\href@noop {} {\bibfield  {journal} {\bibinfo  {journal} {J.
  Phys.: Condens. Mater},\ }\textbf {\bibinfo {volume} {20}},\ \bibinfo {pages}
  {424217} (\bibinfo {year} {2008})}\BibitemShut {NoStop}%
\bibitem [{\citenamefont {Efremov}\ \emph {et~al.}(2004)\citenamefont
  {Efremov}, \citenamefont {Brink},\ and\ \citenamefont
  {Khomskii}}]{efremov_nm}%
  \BibitemOpen
  \bibfield  {author} {\bibinfo {author} {\bibfnamefont {D.~V.}\ \bibnamefont
  {Efremov}}, \bibinfo {author} {\bibfnamefont {J.~V.~D.}\ \bibnamefont
  {Brink}}, \ and\ \bibinfo {author} {\bibfnamefont {D.~I.}\ \bibnamefont
  {Khomskii}},\ }\href@noop {} {\bibfield  {journal} {\bibinfo  {journal} {Nat.
  Mater.},\ }\textbf {\bibinfo {volume} {3}},\ \bibinfo {pages} {853} (\bibinfo
  {year} {2004})}\BibitemShut {NoStop}%
\bibitem [{\citenamefont {Yamauchi}\ \emph {et~al.}(2009)\citenamefont
  {Yamauchi}, \citenamefont {Fukushima},\ and\ \citenamefont
  {Picozzi}}]{yamauchi_prb}%
  \BibitemOpen
  \bibfield  {author} {\bibinfo {author} {\bibfnamefont {K.}~\bibnamefont
  {Yamauchi}}, \bibinfo {author} {\bibfnamefont {T.}~\bibnamefont {Fukushima}},
  \ and\ \bibinfo {author} {\bibfnamefont {S.}~\bibnamefont {Picozzi}},\
  }\href@noop {} {\bibfield  {journal} {\bibinfo  {journal} {Phys. Rev. B},\
  }\textbf {\bibinfo {volume} {79}},\ \bibinfo {pages} {212404} (\bibinfo
  {year} {2009})}\BibitemShut {NoStop}%
\bibitem [{\citenamefont {Fukushima}\ \emph {et~al.}(2011)\citenamefont
  {Fukushima}, \citenamefont {Yamauchi},\ and\ \citenamefont
  {Picozzi}}]{fukushima_jpsj}%
  \BibitemOpen
  \bibfield  {author} {\bibinfo {author} {\bibfnamefont {T.}~\bibnamefont
  {Fukushima}}, \bibinfo {author} {\bibfnamefont {K.}~\bibnamefont {Yamauchi}},
  \ and\ \bibinfo {author} {\bibfnamefont {S.}~\bibnamefont {Picozzi}},\
  }\href@noop {} {\bibfield  {journal} {\bibinfo  {journal} {J. Phys. Soc.
  Jpn.},\ }\textbf {\bibinfo {volume} {80}},\ \bibinfo {pages} {014709}
  (\bibinfo {year} {2011})}\BibitemShut {NoStop}%
\bibitem [{\citenamefont {Wright}\ \emph {et~al.}(2001)\citenamefont {Wright},
  \citenamefont {Attfield},\ and\ \citenamefont {Radaelli}}]{wright_prl}%
  \BibitemOpen
  \bibfield  {author} {\bibinfo {author} {\bibfnamefont {J.~P.}\ \bibnamefont
  {Wright}}, \bibinfo {author} {\bibfnamefont {J.~P.}\ \bibnamefont
  {Attfield}}, \ and\ \bibinfo {author} {\bibfnamefont {P.~G.}\ \bibnamefont
  {Radaelli}},\ }\href@noop {} {\bibfield  {journal} {\bibinfo  {journal}
  {Phys. Rev. Lett.},\ }\textbf {\bibinfo {volume} {87}},\ \bibinfo {pages}
  {266401} (\bibinfo {year} {2001})}\BibitemShut {NoStop}%
\bibitem [{\citenamefont {Blasco}\ \emph {et~al.}(2011)\citenamefont {Blasco},
  \citenamefont {Garcia},\ and\ \citenamefont {Subias}}]{blasco_prb}%
  \BibitemOpen
  \bibfield  {author} {\bibinfo {author} {\bibfnamefont {J.}~\bibnamefont
  {Blasco}}, \bibinfo {author} {\bibfnamefont {J.}~\bibnamefont {Garcia}}, \
  and\ \bibinfo {author} {\bibfnamefont {G.}~\bibnamefont {Subias}},\
  }\href@noop {} {\bibfield  {journal} {\bibinfo  {journal} {Phys. Rev. B},\
  }\textbf {\bibinfo {volume} {83}},\ \bibinfo {pages} {104105} (\bibinfo
  {year} {2011})}\BibitemShut {NoStop}%
\bibitem [{\citenamefont {Joly}\ \emph {et~al.}(2008)\citenamefont {Joly},
  \citenamefont {Lorenzo}, \citenamefont {Nazarenko}, \citenamefont {Hodeau},
  \citenamefont {Mannix},\ and\ \citenamefont {Marin}}]{joly_prb}%
  \BibitemOpen
  \bibfield  {author} {\bibinfo {author} {\bibfnamefont {Y.}~\bibnamefont
  {Joly}}, \bibinfo {author} {\bibfnamefont {J.~E.}\ \bibnamefont {Lorenzo}},
  \bibinfo {author} {\bibfnamefont {E.}~\bibnamefont {Nazarenko}}, \bibinfo
  {author} {\bibfnamefont {J.~L.}\ \bibnamefont {Hodeau}}, \bibinfo {author}
  {\bibfnamefont {D.}~\bibnamefont {Mannix}}, \ and\ \bibinfo {author}
  {\bibfnamefont {C.}~\bibnamefont {Marin}},\ }\href@noop {} {\bibfield
  {journal} {\bibinfo  {journal} {Phys. Rev. B},\ }\textbf {\bibinfo {volume}
  {78}},\ \bibinfo {pages} {134110} (\bibinfo {year} {2008})}\BibitemShut
  {NoStop}%
\bibitem [{\citenamefont {Lorenzo}\ \emph {et~al.}(2008)\citenamefont
  {Lorenzo}, \citenamefont {Mazzoli}, \citenamefont {Jaouen}, \citenamefont
  {Detlefs}, \citenamefont {Mannix}, \citenamefont {Grenier}, \citenamefont
  {Joly},\ and\ \citenamefont {Marin}}]{lorenzo_prl}%
  \BibitemOpen
  \bibfield  {author} {\bibinfo {author} {\bibfnamefont {J.~E.}\ \bibnamefont
  {Lorenzo}}, \bibinfo {author} {\bibfnamefont {C.}~\bibnamefont {Mazzoli}},
  \bibinfo {author} {\bibfnamefont {N.}~\bibnamefont {Jaouen}}, \bibinfo
  {author} {\bibfnamefont {C.}~\bibnamefont {Detlefs}}, \bibinfo {author}
  {\bibfnamefont {D.}~\bibnamefont {Mannix}}, \bibinfo {author} {\bibfnamefont
  {S.}~\bibnamefont {Grenier}}, \bibinfo {author} {\bibfnamefont
  {Y.}~\bibnamefont {Joly}}, \ and\ \bibinfo {author} {\bibfnamefont
  {C.}~\bibnamefont {Marin}},\ }\href@noop {} {\bibfield  {journal} {\bibinfo
  {journal} {Phys. Rev. Lett},\ }\textbf {\bibinfo {volume} {101}},\ \bibinfo
  {pages} {226401} (\bibinfo {year} {2008})}\BibitemShut {NoStop}%
\bibitem [{\citenamefont {Senn}\ \emph
  {et~al.}(2012){\natexlab{a}}\citenamefont {Senn}, \citenamefont {Wright},\
  and\ \citenamefont {Attfield}}]{senn_n}%
  \BibitemOpen
  \bibfield  {author} {\bibinfo {author} {\bibfnamefont {M.~S.}\ \bibnamefont
  {Senn}}, \bibinfo {author} {\bibfnamefont {J.~P.}\ \bibnamefont {Wright}}, \
  and\ \bibinfo {author} {\bibfnamefont {P.~A.}\ \bibnamefont {Attfield}},\
  }\href@noop {} {\bibfield  {journal} {\bibinfo  {journal} {Nature},\ }\textbf
  {\bibinfo {volume} {481}},\ \bibinfo {pages} {173} (\bibinfo {year}
  {2012}{\natexlab{a}})}\BibitemShut {NoStop}%
\bibitem [{\citenamefont {Senn}\ \emph
  {et~al.}(2012){\natexlab{b}}\citenamefont {Senn}, \citenamefont {Loa},
  \citenamefont {Wright},\ and\ \citenamefont {Attfield}}]{senn_prb}%
  \BibitemOpen
  \bibfield  {author} {\bibinfo {author} {\bibfnamefont {M.~S.}\ \bibnamefont
  {Senn}}, \bibinfo {author} {\bibfnamefont {I.}~\bibnamefont {Loa}}, \bibinfo
  {author} {\bibfnamefont {J.~P.}\ \bibnamefont {Wright}}, \ and\ \bibinfo
  {author} {\bibfnamefont {P.~A.}\ \bibnamefont {Attfield}},\ }\href@noop {}
  {\bibfield  {journal} {\bibinfo  {journal} {Phys. Rev. B},\ }\textbf
  {\bibinfo {volume} {85}},\ \bibinfo {pages} {125119} (\bibinfo {year}
  {2012}{\natexlab{b}})}\BibitemShut {NoStop}%
\bibitem [{\citenamefont {Ikeda}\ \emph {et~al.}(2005)\citenamefont {Ikeda},
  \citenamefont {Ohsumi}, \citenamefont {Ohwada}, \citenamefont {Ishii},
  \citenamefont {Inami}, \citenamefont {Kakurai}, \citenamefont {Murakami},
  \citenamefont {Yoshii}, \citenamefont {Mori}, \citenamefont {Horibe},\ and\
  \citenamefont {Kito}}]{ikeda_n}%
  \BibitemOpen
  \bibfield  {author} {\bibinfo {author} {\bibfnamefont {N.}~\bibnamefont
  {Ikeda}}, \bibinfo {author} {\bibfnamefont {H.}~\bibnamefont {Ohsumi}},
  \bibinfo {author} {\bibfnamefont {K.}~\bibnamefont {Ohwada}}, \bibinfo
  {author} {\bibfnamefont {K.}~\bibnamefont {Ishii}}, \bibinfo {author}
  {\bibfnamefont {T.}~\bibnamefont {Inami}}, \bibinfo {author} {\bibfnamefont
  {K.}~\bibnamefont {Kakurai}}, \bibinfo {author} {\bibfnamefont
  {Y.}~\bibnamefont {Murakami}}, \bibinfo {author} {\bibfnamefont
  {K.}~\bibnamefont {Yoshii}}, \bibinfo {author} {\bibfnamefont
  {S.}~\bibnamefont {Mori}}, \bibinfo {author} {\bibfnamefont {Y.}~\bibnamefont
  {Horibe}}, \ and\ \bibinfo {author} {\bibfnamefont {H.}~\bibnamefont
  {Kito}},\ }\href@noop {} {\bibfield  {journal} {\bibinfo  {journal}
  {Nature},\ }\textbf {\bibinfo {volume} {436}},\ \bibinfo {pages} {1136}
  (\bibinfo {year} {2005})}\BibitemShut {NoStop}%
\bibitem [{\citenamefont {Ikeda}\ \emph {et~al.}(2000)\citenamefont {Ikeda},
  \citenamefont {Kohn}, \citenamefont {Myouga}, \citenamefont {Takahashi},
  \citenamefont {Kitoh},\ and\ \citenamefont {Takekawa}}]{ikeda_jpsj}%
  \BibitemOpen
  \bibfield  {author} {\bibinfo {author} {\bibfnamefont {N.}~\bibnamefont
  {Ikeda}}, \bibinfo {author} {\bibfnamefont {K.}~\bibnamefont {Kohn}},
  \bibinfo {author} {\bibfnamefont {N.}~\bibnamefont {Myouga}}, \bibinfo
  {author} {\bibfnamefont {E.}~\bibnamefont {Takahashi}}, \bibinfo {author}
  {\bibfnamefont {H.}~\bibnamefont {Kitoh}}, \ and\ \bibinfo {author}
  {\bibfnamefont {S.}~\bibnamefont {Takekawa}},\ }\href@noop {} {\bibfield
  {journal} {\bibinfo  {journal} {J. Phys. Soc. Jpn.},\ }\textbf {\bibinfo
  {volume} {69}},\ \bibinfo {pages} {1526} (\bibinfo {year}
  {2000})}\BibitemShut {NoStop}%
\bibitem [{\citenamefont {Tokunaga}\ \emph {et~al.}(2006)\citenamefont
  {Tokunaga}, \citenamefont {Lottermoser}, \citenamefont {Lee}, \citenamefont
  {Kumai}, \citenamefont {Uchida}, \citenamefont {Arima},\ and\ \citenamefont
  {Tokura}}]{tokunaga_nm}%
  \BibitemOpen
  \bibfield  {author} {\bibinfo {author} {\bibfnamefont {Y.}~\bibnamefont
  {Tokunaga}}, \bibinfo {author} {\bibfnamefont {T.}~\bibnamefont
  {Lottermoser}}, \bibinfo {author} {\bibfnamefont {Y.}~\bibnamefont {Lee}},
  \bibinfo {author} {\bibfnamefont {R.}~\bibnamefont {Kumai}}, \bibinfo
  {author} {\bibfnamefont {M.}~\bibnamefont {Uchida}}, \bibinfo {author}
  {\bibfnamefont {T.}~\bibnamefont {Arima}}, \ and\ \bibinfo {author}
  {\bibfnamefont {Y.}~\bibnamefont {Tokura}},\ }\href@noop {} {\bibfield
  {journal} {\bibinfo  {journal} {Nat. Mater.},\ }\textbf {\bibinfo {volume}
  {5}},\ \bibinfo {pages} {937} (\bibinfo {year} {2006})}\BibitemShut {NoStop}%
\bibitem [{\citenamefont {Chynoweth}(1956)}]{chynoweth_jap}%
  \BibitemOpen
  \bibfield  {author} {\bibinfo {author} {\bibfnamefont {A.~G.}\ \bibnamefont
  {Chynoweth}},\ }\href@noop {} {\bibfield  {journal} {\bibinfo  {journal} {J.
  Appl. Phys.},\ }\textbf {\bibinfo {volume} {27}},\ \bibinfo {pages} {78}
  (\bibinfo {year} {1956})}\BibitemShut {NoStop}%
\bibitem [{\citenamefont {Lines}\ and\ \citenamefont {Glass}()}]{lines}%
  \BibitemOpen
  \bibfield  {author} {\bibinfo {author} {\bibfnamefont {M.~E.}\ \bibnamefont
  {Lines}}\ and\ \bibinfo {author} {\bibfnamefont {A.~M.}\ \bibnamefont
  {Glass}},\ }\href@noop {} {\emph {\bibinfo {title} {Principles and
  Applications of Ferroelectrics and Related Materials (Oxford University
  Press, New York, 1977)}}}\BibitemShut {NoStop}%
\bibitem [{\citenamefont {Bune}\ \emph {et~al.}(1999)\citenamefont {Bune},
  \citenamefont {Zhu}, \citenamefont {Ducharme}, \citenamefont {Blinov},
  \citenamefont {Fridkin}, \citenamefont {Palto}, \citenamefont {Petukhova},\
  and\ \citenamefont {Yudin}}]{bune_jap}%
  \BibitemOpen
  \bibfield  {author} {\bibinfo {author} {\bibfnamefont {A.~V.}\ \bibnamefont
  {Bune}}, \bibinfo {author} {\bibfnamefont {C.}~\bibnamefont {Zhu}}, \bibinfo
  {author} {\bibfnamefont {S.}~\bibnamefont {Ducharme}}, \bibinfo {author}
  {\bibfnamefont {L.~M.}\ \bibnamefont {Blinov}}, \bibinfo {author}
  {\bibfnamefont {V.~M.}\ \bibnamefont {Fridkin}}, \bibinfo {author}
  {\bibfnamefont {S.~P.}\ \bibnamefont {Palto}}, \bibinfo {author}
  {\bibfnamefont {N.~G.}\ \bibnamefont {Petukhova}}, \ and\ \bibinfo {author}
  {\bibfnamefont {S.~G.}\ \bibnamefont {Yudin}},\ }\href@noop {} {\bibfield
  {journal} {\bibinfo  {journal} {J. Appl. Phys.},\ }\textbf {\bibinfo {volume}
  {85}},\ \bibinfo {pages} {7869} (\bibinfo {year} {1999})}\BibitemShut
  {NoStop}%
\bibitem [{\citenamefont {Takahashi}\ \emph {et~al.}(2009)\citenamefont
  {Takahashi}, \citenamefont {Katayama}, \citenamefont {Dahl}, \citenamefont
  {Grepstad}, \citenamefont {Matsumoto},\ and\ \citenamefont
  {Tybell}}]{takahashi_apl}%
  \BibitemOpen
  \bibfield  {author} {\bibinfo {author} {\bibfnamefont {R.}~\bibnamefont
  {Takahashi}}, \bibinfo {author} {\bibfnamefont {M.}~\bibnamefont {Katayama}},
  \bibinfo {author} {\bibfnamefont {O.}~\bibnamefont {Dahl}}, \bibinfo {author}
  {\bibfnamefont {J.~K.}\ \bibnamefont {Grepstad}}, \bibinfo {author}
  {\bibfnamefont {Y.}~\bibnamefont {Matsumoto}}, \ and\ \bibinfo {author}
  {\bibfnamefont {T.}~\bibnamefont {Tybell}},\ }\href@noop {} {\bibfield
  {journal} {\bibinfo  {journal} {Appl. Phys. Lett.},\ }\textbf {\bibinfo
  {volume} {94}},\ \bibinfo {pages} {232901} (\bibinfo {year}
  {2009})}\BibitemShut {NoStop}%
\bibitem [{\citenamefont {Takahashi}\ \emph
  {et~al.}(2012){\natexlab{a}}\citenamefont {Takahashi}, \citenamefont
  {Tybell},\ and\ \citenamefont {Lippmaa}}]{takahashi_jap}%
  \BibitemOpen
  \bibfield  {author} {\bibinfo {author} {\bibfnamefont {R.}~\bibnamefont
  {Takahashi}}, \bibinfo {author} {\bibfnamefont {T.}~\bibnamefont {Tybell}}, \
  and\ \bibinfo {author} {\bibfnamefont {M.}~\bibnamefont {Lippmaa}},\
  }\href@noop {} {\bibfield  {journal} {\bibinfo  {journal} {J. Appl. Phys.},\
  }\textbf {\bibinfo {volume} {112}},\ \bibinfo {pages} {014111} (\bibinfo
  {year} {2012}{\natexlab{a}})}\BibitemShut {NoStop}%
\bibitem [{\citenamefont {Kawasaki}\ \emph {et~al.}(1994)\citenamefont
  {Kawasaki}, \citenamefont {Takahashi}, \citenamefont {Maeda}, \citenamefont
  {Tsuchiya}, \citenamefont {Shinohara}, \citenamefont {Ishiyama},
  \citenamefont {Yonezawa}, \citenamefont {Yoshimoto},\ and\ \citenamefont
  {Koinuma}}]{kawasaki_s}%
  \BibitemOpen
  \bibfield  {author} {\bibinfo {author} {\bibfnamefont {M.}~\bibnamefont
  {Kawasaki}}, \bibinfo {author} {\bibfnamefont {K.}~\bibnamefont {Takahashi}},
  \bibinfo {author} {\bibfnamefont {T.}~\bibnamefont {Maeda}}, \bibinfo
  {author} {\bibfnamefont {R.}~\bibnamefont {Tsuchiya}}, \bibinfo {author}
  {\bibfnamefont {M.}~\bibnamefont {Shinohara}}, \bibinfo {author}
  {\bibfnamefont {O.}~\bibnamefont {Ishiyama}}, \bibinfo {author}
  {\bibfnamefont {T.}~\bibnamefont {Yonezawa}}, \bibinfo {author}
  {\bibfnamefont {M.}~\bibnamefont {Yoshimoto}}, \ and\ \bibinfo {author}
  {\bibfnamefont {H.}~\bibnamefont {Koinuma}},\ }\href@noop {} {\bibfield
  {journal} {\bibinfo  {journal} {Science},\ }\textbf {\bibinfo {volume}
  {266}},\ \bibinfo {pages} {1540} (\bibinfo {year} {1994})}\BibitemShut
  {NoStop}%
\bibitem [{\citenamefont {Koster}\ \emph {et~al.}(1998)\citenamefont {Koster},
  \citenamefont {Kropman}, \citenamefont {Rijnders}, \citenamefont {Blank},\
  and\ \citenamefont {Rogalla}}]{koster_apl}%
  \BibitemOpen
  \bibfield  {author} {\bibinfo {author} {\bibfnamefont {G.}~\bibnamefont
  {Koster}}, \bibinfo {author} {\bibfnamefont {B.~L.}\ \bibnamefont {Kropman}},
  \bibinfo {author} {\bibfnamefont {G.~J. H.~M.}\ \bibnamefont {Rijnders}},
  \bibinfo {author} {\bibfnamefont {D.~H.~A.}\ \bibnamefont {Blank}}, \ and\
  \bibinfo {author} {\bibfnamefont {H.}~\bibnamefont {Rogalla}},\ }\href@noop
  {} {\bibfield  {journal} {\bibinfo  {journal} {Appl. Phys. Lett.},\ }\textbf
  {\bibinfo {volume} {73}},\ \bibinfo {pages} {2920} (\bibinfo {year}
  {1998})}\BibitemShut {NoStop}%
\bibitem [{\citenamefont {Ohashi}\ \emph {et~al.}(1999)\citenamefont {Ohashi},
  \citenamefont {Lippmaa}, \citenamefont {Nakagawa}, \citenamefont {Nagasawa},
  \citenamefont {Koinuma},\ and\ \citenamefont {Kawasaki}}]{ohashi_rsi}%
  \BibitemOpen
  \bibfield  {author} {\bibinfo {author} {\bibfnamefont {S.}~\bibnamefont
  {Ohashi}}, \bibinfo {author} {\bibfnamefont {M.}~\bibnamefont {Lippmaa}},
  \bibinfo {author} {\bibfnamefont {N.}~\bibnamefont {Nakagawa}}, \bibinfo
  {author} {\bibfnamefont {H.}~\bibnamefont {Nagasawa}}, \bibinfo {author}
  {\bibfnamefont {H.}~\bibnamefont {Koinuma}}, \ and\ \bibinfo {author}
  {\bibfnamefont {M.}~\bibnamefont {Kawasaki}},\ }\href@noop {} {\bibfield
  {journal} {\bibinfo  {journal} {Rev. Sci. Instrum.},\ }\textbf {\bibinfo
  {volume} {70}},\ \bibinfo {pages} {178} (\bibinfo {year} {1999})}\BibitemShut
  {NoStop}%
\bibitem [{\citenamefont {Takahashi}\ \emph
  {et~al.}(2012){\natexlab{b}}\citenamefont {Takahashi}, \citenamefont
  {Misumi},\ and\ \citenamefont {Lippmaa}}]{takahashi_cgd}%
  \BibitemOpen
  \bibfield  {author} {\bibinfo {author} {\bibfnamefont {R.}~\bibnamefont
  {Takahashi}}, \bibinfo {author} {\bibfnamefont {H.}~\bibnamefont {Misumi}}, \
  and\ \bibinfo {author} {\bibfnamefont {M.}~\bibnamefont {Lippmaa}},\
  }\href@noop {} {\bibfield  {journal} {\bibinfo  {journal} {Cryst. Growth
  Des.},\ }\textbf {\bibinfo {volume} {12}},\ \bibinfo {pages} {2679} (\bibinfo
  {year} {2012}{\natexlab{b}})}\BibitemShut {NoStop}%
\bibitem [{JCP(){\natexlab{a}}}]{JCPDS_mag}%
  \BibitemOpen
  \href@noop {} {\emph {\bibinfo {title} {JCPDS Card 19-0629}}}\BibitemShut
  {NoStop}%
\bibitem [{JCP(){\natexlab{b}}}]{JCPDS_STO}%
  \BibitemOpen
  \href@noop {} {\emph {\bibinfo {title} {JCPDS Card 35-0734}}}\BibitemShut
  {NoStop}%
\bibitem [{\citenamefont {Margulies}\ \emph {et~al.}(1996)\citenamefont
  {Margulies}, \citenamefont {Parker}, \citenamefont {Spada}, \citenamefont
  {Goldman}, \citenamefont {Li}, \citenamefont {Sinclair},\ and\ \citenamefont
  {Berkowitz}}]{margulies_prb}%
  \BibitemOpen
  \bibfield  {author} {\bibinfo {author} {\bibfnamefont {D.~T.}\ \bibnamefont
  {Margulies}}, \bibinfo {author} {\bibfnamefont {F.~T.}\ \bibnamefont
  {Parker}}, \bibinfo {author} {\bibfnamefont {F.~E.}\ \bibnamefont {Spada}},
  \bibinfo {author} {\bibfnamefont {R.~S.}\ \bibnamefont {Goldman}}, \bibinfo
  {author} {\bibfnamefont {J.}~\bibnamefont {Li}}, \bibinfo {author}
  {\bibfnamefont {R.}~\bibnamefont {Sinclair}}, \ and\ \bibinfo {author}
  {\bibfnamefont {A.~E.}\ \bibnamefont {Berkowitz}},\ }\href@noop {} {\bibfield
   {journal} {\bibinfo  {journal} {Phys. Rev. B},\ }\textbf {\bibinfo {volume}
  {53}},\ \bibinfo {pages} {9175} (\bibinfo {year} {1996})}\BibitemShut
  {NoStop}%
\bibitem [{\citenamefont {Hibma}\ \emph {et~al.}(1999)\citenamefont {Hibma},
  \citenamefont {Voogt}, \citenamefont {Niesen}, \citenamefont {van~der
  Heijden},\ and\ \citenamefont {de~Jonge}}]{hibma_jap}%
  \BibitemOpen
  \bibfield  {author} {\bibinfo {author} {\bibfnamefont {T.}~\bibnamefont
  {Hibma}}, \bibinfo {author} {\bibfnamefont {F.~C.}\ \bibnamefont {Voogt}},
  \bibinfo {author} {\bibfnamefont {L.}~\bibnamefont {Niesen}}, \bibinfo
  {author} {\bibfnamefont {P.~A.~A.}\ \bibnamefont {van~der Heijden}}, \ and\
  \bibinfo {author} {\bibfnamefont {W.~J.~M.}\ \bibnamefont {de~Jonge}},\
  }\href@noop {} {\bibfield  {journal} {\bibinfo  {journal} {J. Appl. Phys.},\
  }\textbf {\bibinfo {volume} {85}},\ \bibinfo {pages} {5291} (\bibinfo {year}
  {1999})}\BibitemShut {NoStop}%
\bibitem [{\citenamefont {Aragon}\ \emph {et~al.}(1985)\citenamefont {Aragon},
  \citenamefont {Buttrey}, \citenamefont {Shepherd},\ and\ \citenamefont
  {Honig}}]{aragon_prb}%
  \BibitemOpen
  \bibfield  {author} {\bibinfo {author} {\bibfnamefont {R.}~\bibnamefont
  {Aragon}}, \bibinfo {author} {\bibfnamefont {D.~J.}\ \bibnamefont {Buttrey}},
  \bibinfo {author} {\bibfnamefont {J.~P.}\ \bibnamefont {Shepherd}}, \ and\
  \bibinfo {author} {\bibfnamefont {J.~M.}\ \bibnamefont {Honig}},\ }\href@noop
  {} {\bibfield  {journal} {\bibinfo  {journal} {Phys. Rev. B},\ }\textbf
  {\bibinfo {volume} {31}},\ \bibinfo {pages} {430} (\bibinfo {year}
  {1985})}\BibitemShut {NoStop}%
\bibitem [{\citenamefont {Verble}(1974)}]{verble_prb}%
  \BibitemOpen
  \bibfield  {author} {\bibinfo {author} {\bibfnamefont {J.~L.}\ \bibnamefont
  {Verble}},\ }\href@noop {} {\bibfield  {journal} {\bibinfo  {journal} {Phys.
  Rev. B},\ }\textbf {\bibinfo {volume} {9}},\ \bibinfo {pages} {5236}
  (\bibinfo {year} {1974})}\BibitemShut {NoStop}%
\bibitem [{\citenamefont {Phase}\ \emph {et~al.}(2006)\citenamefont {Phase},
  \citenamefont {Tiwari}, \citenamefont {Prakash}, \citenamefont {Dubey},
  \citenamefont {Sathe},\ and\ \citenamefont {Choudhary}}]{phase_jap}%
  \BibitemOpen
  \bibfield  {author} {\bibinfo {author} {\bibfnamefont {D.~M.}\ \bibnamefont
  {Phase}}, \bibinfo {author} {\bibfnamefont {S.}~\bibnamefont {Tiwari}},
  \bibinfo {author} {\bibfnamefont {R.}~\bibnamefont {Prakash}}, \bibinfo
  {author} {\bibfnamefont {A.}~\bibnamefont {Dubey}}, \bibinfo {author}
  {\bibfnamefont {V.~G.}\ \bibnamefont {Sathe}}, \ and\ \bibinfo {author}
  {\bibfnamefont {R.~J.}\ \bibnamefont {Choudhary}},\ }\href@noop {} {\bibfield
   {journal} {\bibinfo  {journal} {J. Appl. Phys.},\ }\textbf {\bibinfo
  {volume} {100}},\ \bibinfo {pages} {123703} (\bibinfo {year}
  {2006})}\BibitemShut {NoStop}%
\bibitem [{\citenamefont {Kovtun}\ and\ \citenamefont
  {Shamyakov}(1973)}]{kovtun_ssc}%
  \BibitemOpen
  \bibfield  {author} {\bibinfo {author} {\bibfnamefont {N.~M.}\ \bibnamefont
  {Kovtun}}\ and\ \bibinfo {author} {\bibfnamefont {A.~A.}\ \bibnamefont
  {Shamyakov}},\ }\href@noop {} {\bibfield  {journal} {\bibinfo  {journal}
  {Solid State Commun.},\ }\textbf {\bibinfo {volume} {13}},\ \bibinfo {pages}
  {1345} (\bibinfo {year} {1973})}\BibitemShut {NoStop}%
\bibitem [{\citenamefont {Yanai}\ \emph {et~al.}(1981)\citenamefont {Yanai},
  \citenamefont {Mizoguchi},\ and\ \citenamefont {Iida}}]{yanai_jpsj}%
  \BibitemOpen
  \bibfield  {author} {\bibinfo {author} {\bibfnamefont {K.}~\bibnamefont
  {Yanai}}, \bibinfo {author} {\bibfnamefont {M.}~\bibnamefont {Mizoguchi}}, \
  and\ \bibinfo {author} {\bibfnamefont {S.}~\bibnamefont {Iida}},\ }\href@noop
  {} {\bibfield  {journal} {\bibinfo  {journal} {J. Phys. Soc. Jpn},\ }\textbf
  {\bibinfo {volume} {50}},\ \bibinfo {pages} {65} (\bibinfo {year}
  {1981})}\BibitemShut {NoStop}%
\bibitem [{\citenamefont {Slack}(1962)}]{slack_pr}%
  \BibitemOpen
  \bibfield  {author} {\bibinfo {author} {\bibfnamefont {G.~A.}\ \bibnamefont
  {Slack}},\ }\href@noop {} {\bibfield  {journal} {\bibinfo  {journal} {Phys.
  Rev.},\ }\textbf {\bibinfo {volume} {126}},\ \bibinfo {pages} {427} (\bibinfo
  {year} {1962})}\BibitemShut {NoStop}%
\bibitem [{\citenamefont {Iwauchi}\ \emph {et~al.}(1980)\citenamefont
  {Iwauchi}, \citenamefont {Kita},\ and\ \citenamefont
  {Koizumi}}]{iwauchi_jpsj}%
  \BibitemOpen
  \bibfield  {author} {\bibinfo {author} {\bibfnamefont {K.}~\bibnamefont
  {Iwauchi}}, \bibinfo {author} {\bibfnamefont {Y.}~\bibnamefont {Kita}}, \
  and\ \bibinfo {author} {\bibfnamefont {N.}~\bibnamefont {Koizumi}},\
  }\href@noop {} {\bibfield  {journal} {\bibinfo  {journal} {J. Phys. Soc.
  Jpn.},\ }\textbf {\bibinfo {volume} {49}},\ \bibinfo {pages} {1328} (\bibinfo
  {year} {1980})}\BibitemShut {NoStop}%
\bibitem [{\citenamefont {Kobayashi}\ \emph {et~al.}(1988)\citenamefont
  {Kobayashi}, \citenamefont {Akishige},\ and\ \citenamefont
  {Sawaguchi}}]{kobayashi_jpsj}%
  \BibitemOpen
  \bibfield  {author} {\bibinfo {author} {\bibfnamefont {M.}~\bibnamefont
  {Kobayashi}}, \bibinfo {author} {\bibfnamefont {Y.}~\bibnamefont {Akishige}},
  \ and\ \bibinfo {author} {\bibfnamefont {E.}~\bibnamefont {Sawaguchi}},\
  }\href@noop {} {\bibfield  {journal} {\bibinfo  {journal} {J. Phys. Soc.
  Jpn.},\ }\textbf {\bibinfo {volume} {57}},\ \bibinfo {pages} {3474} (\bibinfo
  {year} {1988})}\BibitemShut {NoStop}%
\bibitem [{\citenamefont {Fu}\ \emph {et~al.}(2009)\citenamefont {Fu},
  \citenamefont {Taniguchi}, \citenamefont {Itoh}, \citenamefont {Koshihara},
  \citenamefont {Yamamoto},\ and\ \citenamefont {Mori}}]{fu_prl}%
  \BibitemOpen
  \bibfield  {author} {\bibinfo {author} {\bibfnamefont {D.}~\bibnamefont
  {Fu}}, \bibinfo {author} {\bibfnamefont {H.}~\bibnamefont {Taniguchi}},
  \bibinfo {author} {\bibfnamefont {M.}~\bibnamefont {Itoh}}, \bibinfo {author}
  {\bibfnamefont {S.}~\bibnamefont {Koshihara}}, \bibinfo {author}
  {\bibfnamefont {N.}~\bibnamefont {Yamamoto}}, \ and\ \bibinfo {author}
  {\bibfnamefont {S.}~\bibnamefont {Mori}},\ }\href@noop {} {\bibfield
  {journal} {\bibinfo  {journal} {Phys. Rev. Lett.},\ }\textbf {\bibinfo
  {volume} {103}},\ \bibinfo {pages} {207601} (\bibinfo {year}
  {2009})}\BibitemShut {NoStop}%
\bibitem [{\citenamefont {Havriliak}\ and\ \citenamefont
  {Negami}(1967)}]{havriliak_p}%
  \BibitemOpen
  \bibfield  {author} {\bibinfo {author} {\bibfnamefont {S.}~\bibnamefont
  {Havriliak}}\ and\ \bibinfo {author} {\bibfnamefont {S.}~\bibnamefont
  {Negami}},\ }\href@noop {} {\bibfield  {journal} {\bibinfo  {journal}
  {Polymer},\ }\textbf {\bibinfo {volume} {8}},\ \bibinfo {pages} {161}
  (\bibinfo {year} {1967})}\BibitemShut {NoStop}%
\bibitem [{\citenamefont {Watton}\ and\ \citenamefont {Todd}(1991)}]{watton_f}%
  \BibitemOpen
  \bibfield  {author} {\bibinfo {author} {\bibfnamefont {R.}~\bibnamefont
  {Watton}}\ and\ \bibinfo {author} {\bibfnamefont {M.~A.}\ \bibnamefont
  {Todd}},\ }\href@noop {} {\bibfield  {journal} {\bibinfo  {journal}
  {Ferroelectrics},\ }\textbf {\bibinfo {volume} {118}},\ \bibinfo {pages}
  {279} (\bibinfo {year} {1991})}\BibitemShut {NoStop}%
\bibitem [{\citenamefont {Iijima}\ \emph {et~al.}(1986)\citenamefont {Iijima},
  \citenamefont {Tomita}, \citenamefont {Takayama},\ and\ \citenamefont
  {Ueno}}]{iijima_jap}%
  \BibitemOpen
  \bibfield  {author} {\bibinfo {author} {\bibfnamefont {K.}~\bibnamefont
  {Iijima}}, \bibinfo {author} {\bibfnamefont {Y.}~\bibnamefont {Tomita}},
  \bibinfo {author} {\bibfnamefont {R.}~\bibnamefont {Takayama}}, \ and\
  \bibinfo {author} {\bibfnamefont {I.}~\bibnamefont {Ueno}},\ }\href@noop {}
  {\bibfield  {journal} {\bibinfo  {journal} {J. Appl. Phys.},\ }\textbf
  {\bibinfo {volume} {60}},\ \bibinfo {pages} {361} (\bibinfo {year}
  {1986})}\BibitemShut {NoStop}%
\end{thebibliography}%

\end{document}